# Effect of noise on output-only modal identification of beams

[1]Gholamreza Jahangiri[1*] Seyed Rasoul Nabavian[2], Mohammad Reza Davoodi[3], Bahram Navayi Neya[4], Seyedamin Mostafavian[5]

[1] George Mason University, Fairfax, USA.

[1,2,3] Babol Noshirvani University of Technology, Babol, Iran.

[5] Payame Noor University (PNU), Tehran, Iran.

Corresponding author: **Gholamreza Jahangiri**

E-mail: [1]gjahangi@gmu.edu, [2]r.nabavian.h@gmail.com, [3]davoodi@nit.ac.ir, [4]navayi@nit.ac.ir, [5]amin.mostafavian@gmail.com

**Abstract.** In most cases, structural health monitoring depends on determination of the modal parameters of the first few modes of a structure. The data that are used to identify these modes by output-only methods include the structure response together with noise. The presence of noise in the structure response affects the modal parameters. The current study was undertaken to examine the possibility and accuracy of identifying the modal parameters of beams in the presence of noise. For this purpose, the modal parameters of the different modes of single-span beams were obtained using output data with different signal-to-noise ratios. The acceleration signals were obtained by transient analysis and then different powers of noise were generated and added to the signals. The modal parameters of the beams were obtained using output-only methods. Parameters having signal-to-noise ratios of ≥13.98 dB for all modes considered were identified. At a signal-to-noise ratio of -6.02 to 13.98 dB (higher noise level), it was not possible to identify the modal parameters of the first mode of beams, but the parameters of the higher modes were identified with good accuracy.

**Keywords:** Structural identification, Output-only method, Single-span beam, Signal-to-noise ratio.

## 1. Introduction

Structural identification has been applied to structural health monitoring, finite element model updating and damage detection [1,2]. This procedure identifies modal parameters such as the natural frequency, mode shape and damping ratio of a structure. Output-only methods are primarily used for structural identification of structures under actual operating conditions because they have been found to have more advantages than other methods [3-6]. The methods of Peak Picking (PP) [7-8], Frequency Domain Decomposition (FDD) [9-11] and Data-Driven Stochastic Subspace Identification (DD-SSI) [12-16] are used widely for output-only estimation of modal parameters.

In order to properly obtain the modal parameters of a structure using output-only methods, the output data must be of good quality[17]. In practice, the measured outputs are the result of inputs, along with a series of unwanted data that is called "noise" [18]. To identify the modal parameters of a structure using output-only methods, output data should have an appropriate signal-to-noise ratio (SNR). As the signals induced by ambient excitation, which is used for output-only modal identification tests, have very low levels of dynamic response (that is signals with low power), so the noise power should generally be reduced. For this purpose, appropriate data acquisition strategies must be adopted to minimize the level of noise [19-22]. After good data acquisition, techniques such as filtering and averaging can be used to reduce the effect of noise [23-25]. Bonness and Jenkins [26] presented a noise removal technique with which an unlimited amount of unwanted correlated noise can be removed from a set of data by modifying the statistical correlation relations and spectral functions. Adeli and Jiang [27] removed noise from traffic flow data using wavelet packet transform techniques. Jiang et al. [28] developed a Bayesian discrete wavelet packet transform denoising approach based on the integration of Bayesian hypothesis testing and wavelet packet analysis. Al-Ghahtani et al. [29] extended an output-only identification of parameters of a multidimensional system from a record of noisy output measurements by using a multiwavelet denoising technique.

Despite these efforts, uncertainty in the sources and the amount of noise in practice mean it is not possible to eliminate noise completely [24,30]; therefore, identification of the modal parameters of a structure is done using noisy data. Dorvash and Pakzad [31] evaluated the effect of measurement noise on physical contribution ratio (PCR). They observed that the PCR is sensitive to the level of noise in the measured response. Shi et al. [32] proposed a novel output-only method to estimate the structural parameters of a shear-beam building under unknown ground excitation. A three-story shear-beam building was used numerically to demonstrate the proposed technique. Two noise levels of 1% and 5% root mean square (RMS) of noise-free signal were considered. Their result indicated that the estimation errors for stiffness coefficients are 1.41%, for 1% RMS noise and 2.42%, for 5% RMS noise, respectively. De Roeck et al. [33] compared two system identification techniques namely PP and SSI for a 15-storey reinforced concrete shear core building. Their results showed that the SSI technique can detect frequencies that are possibly missed with the PP method and gives a more reasonable modal shape in most cases. Peeters et al. [34] determined the modal parameters of Z-24 bridge via different output-only modal identification methods; two of them were PP and SSI. They resulted that the

---

quality of the extracted mode shapes was higher for the SSI method than for the PP method. Kim and Lynch [35] evaluated two output-only system identification methods namely FDD and SSI for a support-excited frame structure. They concluded that similar accuracy of mode shape estimation was confirmed between the time-domain SSI and frequency-domain FDD and in the case of output-only system identification with a limited number of data points, time-domain SSI was recommended rather than frequency-domain FDD due to the low resolution issues of FDD when estimating natural frequency. Gomaa et al. [36] compared three different techniques of output-only identification methods for extracting modal parameters of a two storey resisting steel frame; two of these methods were FDD and SSI. They resulted that a good agreement in identified natural frequency and mode shape was existed with SSI and FDD. Andersen et al. [37] compared the performance of various output-only modal estimation for identification of Z-24 highway bridge; two of these methods were PP and SSI. The two methods seemed to agree very well on the natural frequency and mode shape estimations of the first five modes; except of the first mode that PP couldn't identify it. Yi and Yun [38] investigated several modal identifications for a two bay and 4-storey building structure. They also considered o% to 60 % noise in the root mean squares (RMS) of noise-free signals. Their numerical investigation showed that the frequency domain method (FDD) was generally more vulnerable to the measurement noise than the time domain method (SSI), and the estimates by the frequency domain method were less accurate than those by time domain method particularly when the two adjacent modes were closely spaced. More ever they included that the SSI method gave most accurate estimates under the large measurement noise. Chen et al. [39] obtained the modal parameters of Newmarket Viaduct bridge via different output-only methods; two of them were PP and SSI. The comparison revealed that the two methods considered gave reasonably consistent estimates of the natural frequencies and mode shapes.

These parameters which are obtained with noisy data are then used for structural health monitoring, finite element model updating and damage detection. In most cases, the first few modes of a structure are used [40]. In the current research, it is hypothesized that the possibility and accuracy of identifying the modal parameters of the first few modes of a structure are affected by the SNR.

The effects of various SNRs on output-only identification of modal parameters were studied. For this purpose, four beams with the same geometrical and mechanical properties, but different support conditions were considered. Band-limited white noise excitation was applied to each beam and the acceleration signals from the beam were obtained using transient analysis. Using these signals, the modal parameters of the first five modes of the beams were identified using Peak Picking, Frequency Domain Decomposition and Data-Driven Stochastic Subspace Identification methods. Because less attention is paid to damping in structural health monitoring compared to the other two modal parameters (mode shape and natural frequency) [41], it was not identified. To generate noisy data, noise having powers that differed from the signal power were added to the signals. Because of the random nature of noise, this process was repeated 100 times. The noisy data and described methods were then used to obtain the modal parameters of the beams for various SNRs.

## 2. Noisy data generation

Output-only structural identification is based only on the measured response of structures. This response contains the structure response and the output noise. The simplest mathematical model for considering noise in data is the additive noise model shown in Fig. 1 [42]. In this model, the output noise is modeled as zero-mean Gaussian white noise [43]. The structure response $S(t)$, referred to herein as the signal, is corrupted by random output noise $N(t)$. The measured response is $NS(t)$. The output noise consists of any undesirable signal and physically may arise from measurement devices and sensors, etc. [44].

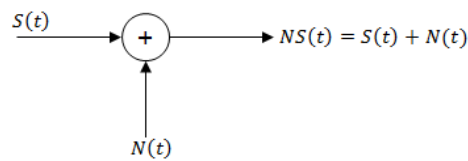

**Fig. 1.** The additive noise model [42].

The amount of noise in the measured response affects the accuracy of the estimated modal parameters [45,46]. The amount of noise in a signal is quantified by the SNR and is expressed in Eq. (1) as:

$$SNR = \frac{P_s}{P_n} \quad (1)$$

where $P_s$ and $P_n$ are the signal power and noise power, respectively, and P is the mean square of data shown in Eq. (2) as:

$$P = \frac{\sum_{i=1}^{n} x_i^2}{n} \quad (2)$$

where $x_i$ denotes the $i^{th}$ data sample in which $n$ is the total number of samples. SNR may be expressed in decibels as:

$$SNR_{dB} = 10 \times \log_{10}(SNR) \quad (3)$$

Eq. (4) is used to generate noise $N(t)$ [47,48] as:

$$N(t) = RMS(S(t)) \times NL \times W(t) \quad (4)$$

where $RMS(S(t))$ is the root mean square (RMS) of the signals, $NL$ is a constant that determines the noise level and $W(t)$ is a random function with the same dimensions as $S(t)$. The presence of $W(t)$ in Eq. (4) is due to the random nature of noise. This function has a normal distribution with a zero mean and unit standard deviation [43]. N(t) is defined as such to allow parameterization of its power with respect to signal power, as will be explained below.

Eq. (4) can be rewritten in Eq. (5) as:

$$\begin{Bmatrix} n_1 \\ \vdots \\ n_n \end{Bmatrix} = RMS\left(\begin{Bmatrix} s_1 \\ \vdots \\ s_n \end{Bmatrix}\right) \times NL \times \begin{Bmatrix} w_1 \\ \vdots \\ w_n \end{Bmatrix} \quad (5)$$

where $n_i$, $s_i$ and $w_i$ are the noise, signal and random function of the $i^{th}$ data sample, respectively. Eq. (5) leads to the formation of Eq. (6) as:

$$\begin{cases} n_1 = RMS(S(t)) \times NL \times w_1 \\ \qquad \vdots \\ n_n = RMS(S(t)) \times NL \times w_n \end{cases} \quad (6)$$

Squaring both sides of Eq. (6) and then taking the average produces:

$$\frac{\sum_{i=1}^{n} n_i^2}{n} = [RMS(S(t))]^2 \times (NL)^2 \times \frac{\sum_{i=1}^{n} w_i^2}{n} \quad (7)$$

Given $RMS(S(t)) = \sqrt{P_s}$, $P_n = \sum_{i=1}^{n} n_i^2 / n$, Eq. (7) can be rewritten as follows:

$$P_n = P_s \times (NL)^2 \times \frac{\sum_{i=1}^{n} w_i^2}{n} \quad (8)$$

Because $W(t)$ has zero mean and unit standard deviation, $\frac{\sum_{i=1}^{n} w_i^2}{n}$ in Eq. (8) is equal to the variance of random function $W(t)$, which is equal to one. Eq. (9) then can be rewritten as follows:

$$P_n = P_s \times (NL)^2 \quad (9)$$

In Eq. (9), the noise power equals the product of the signal power and the square of the noise level. The SNR in this case is equal to the inverse of the square of the noise level as shown in Eq. (10):

$$SNR = \frac{P_s}{P_n} = \frac{1}{(NL)^2} \quad (10)$$

In term of decibels, Eq. (10) may be expressed in Eq. (11) as:

$$SNR_{dB} = 10 \times \log_{10} \frac{1}{(NL)^2} = -20 \times \log_{10} NL \quad (11)$$

After generating the noise and adding it to the signal, a noisy signal can be generated in Eq. (12) [49] as:

$$NS(t) = S(t) + N(t) \quad (12)$$

## 3. Modeling and generating noise-free data

To investigate the effect of SNR on the structural identification of single-span beams, four beams having different support conditions were considered. In Fig. 2, these beams are shown as a clamped-free supported beam (CF), simple-simple supported beam (SS), clamped-simple supported beam (CS) and clamped-clamped supported beam (CC). ANSYS finite element analysis software was used for analysis of the beams. Of the library elements of this software, the element BEAM 188 was used for modeling of the beams. BEAM 188 is a linear (2-node) beam element with six degrees of freedom at each node. The degrees of freedom at each node include translations in and rotations about the x, y, and z directions [50].

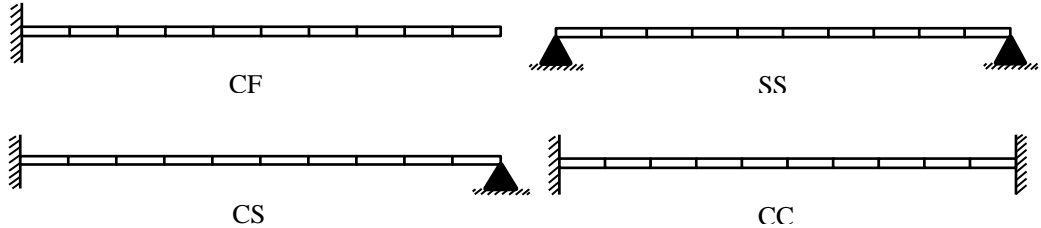

**Fig. 2.** Finite element model of beams CF, SS, CS and CC.

Each beam was modeled in two-dimensional space with ten elements, each with a length of 100 mm and a square cross-section of $10 \times 10$ mm. Sensitivity analysis was used to determine that an element length of 100 mm provides acceptable accuracy for studying the modes of interest in this work. The material model behavior of the beams is linear isotropic. The elastic moduli of the beams is $2\times10^{11}$ N/m$^2$, the mass density is 7850 N.s$^2$/m$^4$ and the Poisson's ratio is 0.3. These are typical properties of steel sections produced in Iran. The geometric properties of the beams were chosen such that the frequencies of the modes considered are sufficiently far apart.

For a validation of the finite element model in ANSYS, the analytical natural frequencies of the beam CF was obtained via dynamic equation of a clamped-free beam subjected to free vibration [51]. The results for the first five natural frequency of the beam CF with the relative error percentage between modal analysis of finite element model and analytical equation are presented in Table 1. As can be seen, the difference of the natural frequencies in higher modes is greater due to the limitation of the meshing of the finite element model.

**Table 1.** The first five natural frequencies of the beams CF obtained from the modal analysis and analytical method.

| Mode Number | Natural Frequency [Hz] | Relative Error [%] |
| --- | --- | --- |

|       | Modal Analysis | Analytical method |      |
|-------|----------------|-------------------|------|
| 1st Mode | 8.2         | 8.2               | 0.0  |
| 2nd Mode | 52.2        | 51.2              | 1.9  |
| 3rd Mode | 152.4       | 144.8             | 4.9  |
| 4th Mode | 319.0       | 280.8             | 12.0 |
| 5th Mode | 577.3       | 463.7             | 19.7 |

The modal parameters for each beam were obtained using finite element modal analysis and are considered to be reference modal parameters. The first five natural frequencies of the beams SS, CS and CC are shown in Table 2.

**Table 2.** The first five natural frequencies of the beams SS, CS and CC obtained from the modal analysis.

| Beam | Natural Frequency [Hz] | | | | |
|------|--------|--------|--------|--------|--------|
|      | Mode 1 | Mode 2 | Mode 3 | Mode 4 | Mode 5 |
| SS   | 23.2   | 95.9   | 228.7  | 442.5  | 773.8  |
| CS   | 36.4   | 123.0  | 274.0  | 513.9  | 885.3  |
| CC   | 53.3   | 154.1  | 325.0  | 594.1  | 1011.2 |

To generate noise-free data from transient analysis of beams, each node was excited in the vertical direction using a time history force which was generated as Gaussian-banded white noise (1 to 1500 Hz) with RMS amplitudes of 0.2 Newton. The sampling time step was 0.0001 sec and the acceleration responses were measured at a rate of 10000 Hz, resulting in a Nyquist frequency of 5000 Hz, which is higher than the largest natural frequency of all the beams in Table 1 (<1011.2 Hz).

The recording duration (T) was the required time for a certain mode and can be defined as $(\omega_k \xi_k)^{-1}$ where $\omega_k$ and $\xi_k$ are the natural frequency and damping ratio of mode k, respectively [50]. Proportional damping was assigned to the model by matching a damping ratio of 2.5% to all frequencies. As the minimum natural frequency of the beams was 8.2 Hz in Table 1, the recording duration based on the above formula was 4.9 sec. A recording duration of 5 sec was considered in this work.

**4. Structural identification of beams with noisy data**

In Section 3, the acceleration signals were obtained from transient analysis of each beam. These signals (S(t)) were free of noise (zero noise level). Measurement noise N(t) was modeled as zero-mean Gaussian white noise and was added to all channels of acceleration response data. The noise level was defined as the ratio of the RMS of the noise to the RMS of the noise-free acceleration response at each channel. This ratio was kept constant at all channels for a given noise level. Seven levels of measurement noise were considered here. Noise N(t) was generated at different noise levels (NL) as shown in Eq. (4). By adding N(t) to S(t), as in Eq. (12), noisy data (NS(t)) was generated. The noise levels and the corresponding SNR values are shown in Table 3.

**Table 3.** Different noise levels and $SNR_{dB}$ values.

| Noise Level (NL) | signal to noise ratio (dB) |
|------------------|----------------------------|
| 5%               | 26.02                      |
| 10%              | 20.00                      |
| 20%              | 13.98                      |
| 50%              | 6.02                       |
| 75%              | 2.50                       |
| 100%             | 0.00                       |
| 200%             | -6.02                      |

Noise generation is statistically independent; thus, because of the random characteristics of the noise, the process should be repeated [33]. A set of 100 system identification runs was performed for each beam. At each noise level, 100 noise N(t) and, consequently, 100 noisy data sets NS(t) were generated. The first five mode shapes of each beam were determined with the noisy data NS(t) using the PP method. The modal assurance criterion (MAC) values were calculated between each of these mode shapes along with the corresponding reference mode shapes (MAC values of zero to one; where one represents full compliance of the two modes [53]). The minimum mean and standard deviation of MAC for 100 runs of beam CF are shown in Table 4. Because the minimum values of MAC for all mode shapes of the beam at SNRdB = 26.02 are equal to 1, these mode shapes as identified by PP were in good agreement with the reference mode shapes. Table 4 shows that, for the first mode shape, decreasing SNRdB decreased the minimum and mean values of MAC and increased the standard deviations. No significant change occurred in the minimum, mean and standard deviation of MAC in other mode shapes after decreasing SNRdB. This process was performed for the SS, CS and CC beams. The noisy data with the minimum MAC value for each beam using the PP method was used to determine the modal parameters for the other methods.

The frequency resolution for PP and FDD was 1/T = 1/5 Hz = 0.2 Hz. The peaks were selected manually in PP and FDD. For implementation of the SSI data-driven method, the acceleration response data was used to form an output Hankel matrix having 10 block rows with 9 or 10 rows in each block (equal to the number of acceleration channels considered). The number of block rows multiplied by the number of measurement channels will produce the maximum model order. For noisy data, several model orders were considered to build block Hankel matrix which was at most 200.

**Table 4.** Minimum, mean and standard deviation of MAC values of 100 runs for beam CF.

| Mode No. | Statistical Result | SNR (dB) | | | | | | |
|---|---|---|---|---|---|---|---|---|
| | | 26.02 | 20.00 | 13.98 | 6.02 | 2.50 | 0.00 | -6.02 |
| 1st Mode | Minimum | 1.00 | 0.99 | 0.98 | 0.88 | 0.77 | 0.59 | 0.00 |
| | Average | 1.00 | 1.00 | 0.99 | 0.96 | 0.90 | 0.82 | 0.50 |
| | Standard Deviation | 0.00 | 0.01 | 0.03 | 0.15 | 0.31 | 0.64 | 2.17 |
| 2nd Mode | Minimum | 1.00 | 1.00 | 1.00 | 1.00 | 0.99 | 0.99 | 0.95 |
| | Average | 1.00 | 1.00 | 1.00 | 1.00 | 1.00 | 1.00 | 0.98 |
| | Standard Deviation | 0.00 | 0.00 | 0.00 | 0.00 | 0.01 | 0.02 | 0.03 |
| 3rd Mode | Minimum | 1.00 | 1.00 | 1.00 | 1.00 | 1.00 | 1.00 | 0.99 |
| | Average | 1.00 | 1.00 | 1.00 | 1.00 | 1.00 | 1.00 | 1.00 |
| | Standard Deviation | 0.00 | 0.00 | 0.00 | 0.00 | 0.00 | 0.00 | 0.01 |
| 4th Mode | Minimum | 1.00 | 1.00 | 1.00 | 1.00 | 1.00 | 1.00 | 1.00 |
| | Average | 1.00 | 1.00 | 1.00 | 1.00 | 1.00 | 1.00 | 1.00 |
| | Standard Deviation | 0.00 | 0.00 | 0.00 | 0.00 | 0.00 | 0.00 | 0.00 |
| 5th Mode | Minimum | 1.00 | 1.00 | 1.00 | 1.00 | 1.00 | 1.00 | 0.99 |
| | Average | 1.00 | 1.00 | 1.00 | 1.00 | 1.00 | 1.00 | 1.00 |
| | Standard Deviation | 0.00 | 0.00 | 0.00 | 0.00 | 0.00 | 0.00 | 0.01 |

### 4.1. Identification of natural frequencies

The first five natural frequencies of each beam were determined using the PP, FDD and SSI methods and the results are presented in Tables 4-7. The second column of these tables shows the SNRdB values for the various noise levels. The differences between these values and the corresponding values shown in Table 3 are due to the random characteristics of W(t) in Eq. (3). The MATLAB function randn was used to generate random function W(t). Because the random numbers generated have no exact zero means and unit standard deviations, the resulting SNRdB values may differ slightly from values in Table 3. For example, for beam CF, the SNRdB for a noise level of 5% was 26.08 instead of 26.02, which is a difference of 0.23%. The natural frequencies of noise-free beams are presented in row NL = 0 in Tables 5 to 8. In this case, the SNRdB value is denoted by an infinity symbol. In Tables 5 to 8, some natural frequencies were not identified (as denoted by a dash). It can be seen that FDD performed better for noisy data than the other methods because it obtained more natural frequencies. For example, the first natural frequency of beam CF was obtained at a noise level of 50%, while such a frequency was identified using PP and SSI at 20% and 5%, respectively.

**Table 5**: The identified natural frequencies of beam CF (Hz)

| Noise Level (NL) | SNR$_{dB}$ | Frequency of Mode 1 | | | Frequency of Mode 2 | | | Frequency of Mode 3 | | | Frequency of Mode 4 | | | Frequency of Mode 5 | | |
|---|---|---|---|---|---|---|---|---|---|---|---|---|---|---|---|---|
| | | PP | FDD | SSI | PP | FDD | SSI | PP | FDD | SSI | PP | FDD | SSI | PP | FDD | SSI |
| 0 | ∞ | 8.0 | 7.3 | 8.1 | 52.0 | 51.3 | 52.1 | 152.0 | 151.4 | 152.2 | 318.0 | 317.4 | 317.8 | 571.0 | 571.3 | 571.1 |
| 5% | 26.08 | 8.0 | 7.3 | 8.3 | 52.0 | 51.3 | 52.2 | 152.0 | 151.4 | 152.2 | 318.0 | 317.4 | 317.1 | 571.0 | 571.3 | 571.1 |
| 10% | 20.10 | 8.0 | 7.3 | - | 52.0 | 51.3 | 52.0 | 152.0 | 151.4 | 152.2 | 318.0 | 317.4 | 317.8 | 571.0 | 571.3 | 570.9 |
| 20% | 13.95 | 8.0 | 7.3 | - | 52.0 | 51.3 | 52.4 | 152.0 | 151.4 | 152.2 | 318.0 | 317.4 | 317.8 | 571.0 | 571.3 | 571.1 |
| 50% | 6.05 | - | 7.3 | - | 52.0 | 51.3 | 55.8 | 152.0 | 151.4 | 152.3 | 318.0 | 317.4 | 317.8 | 571.0 | 571.3 | 571.1 |
| 75% | 2.46 | - | - | - | 52.0 | 51.3 | 54.1 | 152.0 | 151.4 | 152.5 | 318.0 | 317.4 | 317.8 | 571.0 | 571.3 | 571.0 |
| 100% | 0.04 | - | - | - | 52.0 | 51.3 | 55.4 | 152.0 | 151.4 | 152.5 | 318.0 | 317.4 | 318.2 | 571.0 | 571.3 | 571.3 |
| 200% | -6.02 | - | - | - | 52.0 | 51.3 | 55.7 | 152.0 | 151.4 | 150.9 | 318.0 | 317.4 | 317.4 | 571.0 | 571.3 | 571.5 |

**Table 6:** The identified natural frequencies of beam SS (Hz)

| Noise Level (NL) | SNR$_{dB}$ | Frequency of Mode 1 | | | Frequency of Mode 2 | | | Frequency of Mode 3 | | | Frequency of Mode 4 | | | Frequency of Mode 5 | | |
|---|---|---|---|---|---|---|---|---|---|---|---|---|---|---|---|---|
| | | PP | FDD | SSI | PP | FDD | SSI | PP | FDD | SSI | PP | FDD | SSI | PP | FDD | SSI |
| 0 | ∞ | 23.0 | 22.0 | 23.6 | 96.0 | 95.2 | 96.0 | 228.5 | 229.5 | 228.2 | 439.5 | 439.5 | 439.2 | 759.0 | 759.3 | 758.9 |
| %5 | 25.96 | 23.0 | 22.0 | 23.6 | 96.0 | 95.2 | 96.0 | 228.5 | 229.5 | 228.2 | 439.5 | 439.5 | 439.2 | 759.0 | 759.3 | 758.9 |
| %10 | 20.13 | 23.0 | 22.0 | 23.9 | 96.0 | 95.2 | 96.0 | 228.5 | 229.5 | 228.5 | 439.5 | 439.5 | 439.5 | 759.0 | 759.3 | 758.8 |
| %20 | 13.88 | 23.0 | 22.0 | 23.3 | 96.0 | 95.2 | 96.1 | 228.5 | 229.5 | 228.5 | 439.5 | 439.5 | 439.4 | 759.0 | 759.3 | 757.8 |
| %50 | 5.93 | 23.0 | 22.0 | - | 96.0 | 95.2 | 96.6 | 228.5 | 229.5 | 228.8 | 439.5 | 439.5 | 439.4 | 759.0 | 759.3 | 759.1 |
| %75 | 2.48 | 23.0 | 22.0 | - | 96.0 | 95.2 | 96.6 | 228.5 | 229.5 | 228.8 | 439.5 | 439.5 | 439.5 | 759.0 | 759.3 | 759.4 |
| %100 | 0.09 | - | - | - | 96.0 | 95.2 | 96.0 | 228.5 | 229.5 | 228.8 | 439.5 | 439.5 | 439.4 | 759.0 | 759.3 | 759.3 |
| %200 | -6.02 | - | - | - | 96.0 | 95.2 | 96.1 | 228.5 | 229.5 | 229.0 | 439.5 | 439.5 | 439.5 | 759.0 | 759.3 | 759.0 |

**Table 7:** The identified natural frequencies of beam CS (Hz)

| Noise Level (NL) | SNR$_{dB}$ | Frequency of Mode 1 | | | Frequency of Mode 2 | | | Frequency of Mode 3 | | | Frequency of Mode 4 | | | Frequency of Mode 5 | | |
|---|---|---|---|---|---|---|---|---|---|---|---|---|---|---|---|---|
| | | PP | FDD | SSI | PP | FDD | SSI | PP | FDD | SSI | PP | FDD | SSI | PP | FDD | SSI |
| 0 | ∞ | 36.5 | 36.6 | 36.4 | 123.0 | 122.1 | 122.9 | 273.5 | 273.4 | 273.3 | 509.5 | 510.3 | 509.7 | 863.5 | 864.3 | 863.8 |
| %5 | 26.09 | 36.5 | 36.6 | 36.4 | 123.0 | 122.1 | 122.9 | 273.5 | 273.4 | 273.3 | 509.5 | 510.3 | 509.7 | 863.5 | 864.3 | 863.8 |
| %10 | 20.09 | 36.5 | 36.6 | 36.3 | 123.0 | 122.1 | 122.9 | 273.5 | 273.4 | 273.3 | 509.5 | 510.3 | 509.7 | 863.5 | 864.3 | 863.5 |
| %20 | 13.98 | 36.5 | 36.6 | 36.7 | 123.0 | 122.1 | 122.9 | 273.5 | 273.4 | 273.3 | 509.5 | 510.3 | 509.7 | 863.5 | 864.3 | 863.6 |
| %50 | 6.14 | 36.5 | 36.6 | - | 123.0 | 122.1 | 123.0 | 273.5 | 273.4 | 273.3 | 509.5 | 510.3 | 510.4 | 863.5 | 864.3 | 863.7 |
| %75 | 2.50 | 36.5 | 36.6 | - | 123.0 | 122.1 | 122.8 | 273.5 | 273.4 | 273.3 | 509.5 | 510.3 | 510.3 | 863.5 | 864.3 | 863.6 |
| %100 | -0.09 | - | - | - | 123.0 | 122.1 | 122.7 | 273.5 | 273.4 | 273.3 | 509.5 | 510.3 | 510.4 | 863.5 | 864.3 | 864.0 |
| %200 | -6.02 | - | - | - | 123.0 | 122.1 | 123.8 | 273.5 | 273.4 | 274.0 | 509.5 | 510.3 | 510.2 | 863.5 | 864.3 | 864.0 |

**Table 8:** The identified natural frequencies of beam CC (Hz)

| Noise Level (NL) | SNR$_{dB}$ | Frequency of Mode 1 | | | Frequency of Mode 2 | | | Frequency of Mode 3 | | | Frequency of Mode 4 | | | Frequency of Mode 5 | | |
|---|---|---|---|---|---|---|---|---|---|---|---|---|---|---|---|---|
| | | PP | FDD | SSI | PP | FDD | SSI | PP | FDD | SSI | PP | FDD | SSI | PP | FDD | SSI |
| 0 | ∞ | 53.5 | 53.7 | 53.3 | 154.0 | 153.8 | 154.0 | 324.0 | 324.7 | 323.9 | 587.5 | 588.4 | 587.3 | 979.0 | 979.0 | 979.1 |
| %5 | 26.04 | 53.5 | 53.7 | 53.3 | 154.0 | 153.8 | 154.0 | 324.0 | 324.7 | 323.9 | 587.5 | 588.4 | 587.3 | 979.0 | 979.0 | 979.1 |
| %10 | 20.01 | 53.5 | 53.7 | 53.1 | 154.0 | 153.8 | 154.0 | 324.0 | 324.7 | 323.9 | 587.5 | 588.4 | 587.4 | 979.0 | 979.0 | 979.1 |
| %20 | 13.92 | 53.5 | 53.7 | 53.7 | 154.0 | 153.8 | 154.1 | 324.0 | 324.7 | 323.9 | 587.5 | 588.4 | 587.4 | 979.0 | 979.0 | 978.1 |
| %50 | 5.98 | 53.5 | 53.7 | 53.8 | 154.0 | 153.8 | 154.1 | 324.0 | 324.7 | 324.0 | 587.5 | 588.4 | 587.5 | 979.0 | 979.0 | 978.8 |
| %75 | 2.46 | 53.5 | 53.7 | 52.9 | 154.0 | 153.8 | 154.1 | 324.0 | 324.7 | 324.0 | 587.5 | 588.4 | 587.4 | 979.0 | 979.0 | 979.0 |
| %100 | 0.09 | - | 53.7 | - | 154.0 | 153.8 | 153.8 | 324.0 | 324.7 | 323.9 | 587.5 | 588.4 | 587.1 | 979.0 | 979.0 | 978.7 |
| %200 | -6.02 | - | - | - | 154.0 | 153.8 | 153.8 | 324.0 | 324.7 | 323.8 | 587.5 | 588.4 | 589.4 | 979.0 | 979.0 | 979.0 |

It can be seen from Tables 5 to 8 that the natural frequency of the first mode of the beams was identified using PP, FDD and SSI methods for the SNRs as ≥13.98, ≥6.02, and ≥26.02, respectively. The second to fifth natural frequencies of all beams for the SNRs were ≥-6.02.

As shown in Tables 5 to 8, increasing the noise level did not alter the natural frequencies identified using FDD and PP. Gkoktsi et al. [54] obtained the modal parameters of a simply-supported beam using FDD for SNRs of 10, 20 and 10^20 dB. They found that the noise level did not significantly affect the natural frequency estimation in these SNRs. In the frequency domain methods (PP and FDD), the natural frequency of the structure was obtained from the peak value of the ANPSD diagram. The ANPSD diagram for beam CF is shown in Fig. 3 for NL = 0% to 200%. The last free node (node 11) of the beam was used as the reference channel. Fig. 3 shows that, although the different noise levels produced different ANPSD diagrams, the peak location or values of the natural frequencies of the beams did not change.

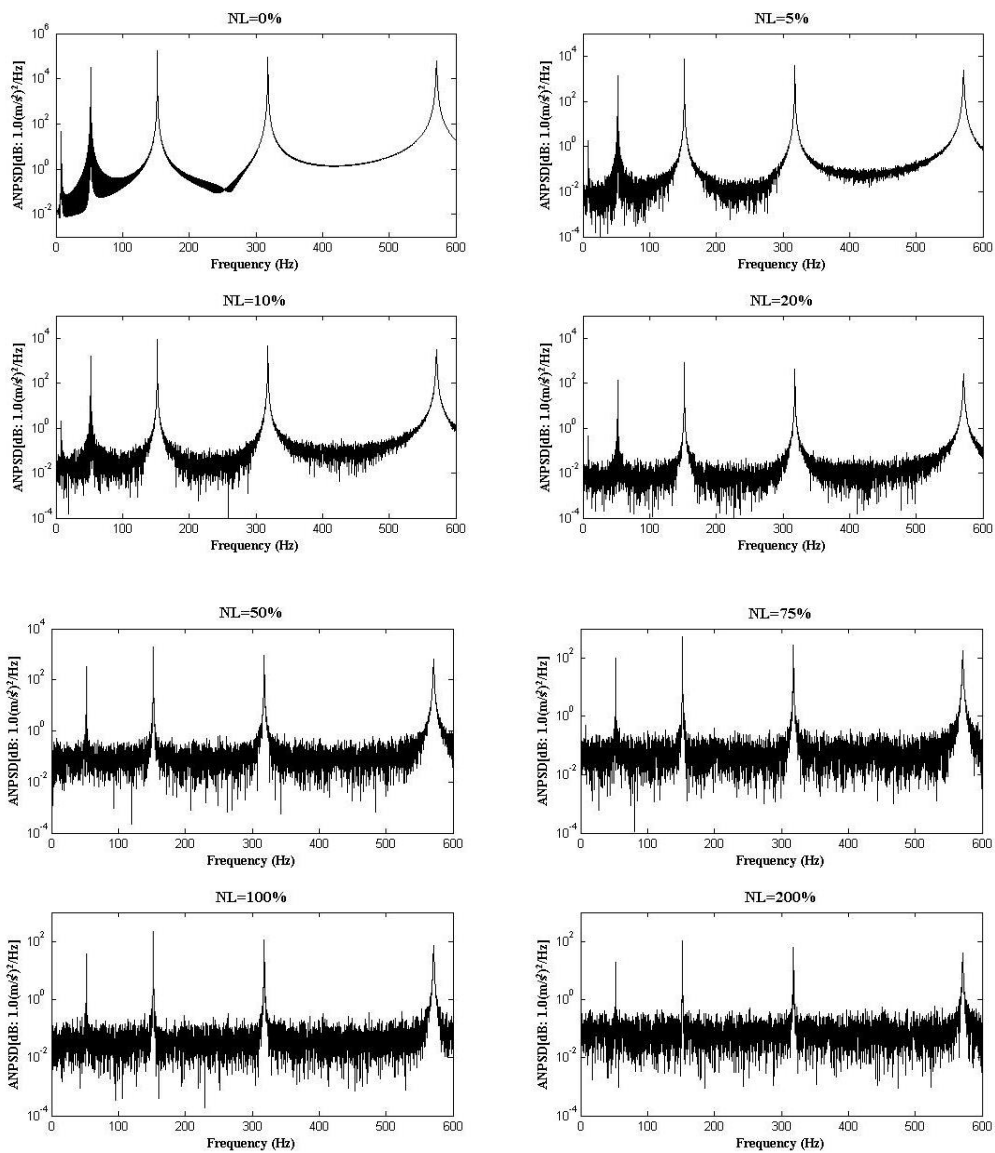

**Fig. 3.** The ANPSD diagram for beam CF for various noise levels

The percentage of relative error between the identified natural frequencies (Tables 5 to 8) and the reference natural frequencies of the beams (Table 2) were determined and are presented in Table 9. For the SSI method, the mean percentage of relative error for various noise levels is presented. Other than the error of the natural frequency of the first mode of beams CF and SS identified by FDD, the errors are very small.

**Table 9:** The relative error between the identified and reference natural frequency for the beams (%).

| Beam | Method | Frequency of Mode 1 | Frequency of Mode 2 | Frequency of Mode 3 | Frequency of Mode 4 | Frequency of Mode 5 |
|---|---|---|---|---|---|---|
| CF | PP | 2.4 | 0.4 | 0.3 | 0.3 | 1.1 |
|  | FDD | 10.9 | 1.7 | 0.7 | 0.5 | 1.0 |
|  | SSI | 1.2 | 3.0 | 0.2 | 0.4 | 1.1 |
| SS | PP | 0.9 | 0.1 | 0.1 | 0.7 | 1.9 |
|  | FDD | 5.2 | 0.7 | 0.3 | 0.7 | 1.9 |
|  | SSI | 1.3 | 0.3 | 0.1 | 0.7 | 1.9 |
| CS | PP | 0.3 | 0.0 | 0.2 | 0.8 | 2.4 |
|  | FDD | 0.5 | 0.7 | 0.2 | 0.7 | 2.4 |
|  | SSI | 0.3 | 0.2 | 0.2 | 0.8 | 2.4 |
| CC | PP | 0.4 | 0.1 | 0.3 | 1.1 | 3.2 |

| | | | Beam CF | | | Beam SS | | | Beam CS | | | Beam CC | | |
|---|---|---|---|---|---|---|---|---|---|---|---|---|---|---|
| | | | PP | FDD | SSI | PP | FDD | SSI | PP | FDD | SSI | PP | FDD | SSI |
| FDD | | | 0.8 | | | 0.5 | | | 0.1 | | | 1.0 | | 3.2 |
| SSI | | | 0.5 | | | 0.1 | | | 0.3 | | | 1.1 | | 3.2 |

### 4.2. Identification of mode shapes

As stated, the first five mode shapes of each beam were identified using the noisy data having a minimum MAC value using the PP method. The MAC values between these mode shapes and the corresponding reference mode shapes are presented in Table 10. A value of 1 in the NL = 0% row indicates good agreement between the identified mode shapes and the corresponding reference mode shapes for the noise-free case. As shown in Table 10, as the noise level increased, the MAC values either decreased or remained constant. The lower values for MAC for the first mode shape of each beam show that this mode of the beams was not identified appropriately at high noise levels. Because the MAC values close or equal to the one obtained by FDD are greater than for the other two methods, this method is more appropriate for the identification of the mode shapes of the beams in the case of noisy data.

**Table 10:** The MAC values between the identified mode shapes and the corresponding reference mode shapes.

| Mode Number | NL | $SNR_{dB}$ | Beam CF | | | Beam SS | | | Beam CS | | | Beam CC | | |
|---|---|---|---|---|---|---|---|---|---|---|---|---|---|---|
| | | | PP | FDD | SSI | PP | FDD | SSI | PP | FDD | SSI | PP | FDD | SSI |
| 1st Mode | 0% | ∞ | 1.00 | 1.00 | 1.00 | 1.00 | 1.00 | 1.00 | 1.00 | 1.00 | 1.00 | 1.00 | 1.00 | 1.00 |
| | 5% | 26.02 | 1.00 | 1.00 | 0.98 | 1.00 | 1.00 | 1.00 | 1.00 | 1.00 | 1.00 | 1.00 | 1.00 | 1.00 |
| | 10% | 20.00 | 0.99 | 1.00 | 0.86 | 1.00 | 1.00 | 0.99 | 1.00 | 1.00 | 0.99 | 1.00 | 1.00 | 1.00 |
| | 20% | 13.98 | 0.98 | 1.00 | 0.84 | 1.00 | 1.00 | 0.97 | 1.00 | 1.00 | 0.96 | 1.00 | 1.00 | 0.99 |
| | 50% | 6.02 | 0.88 | 0.99 | 0.65 | 0.99 | 0.99 | 0.88 | 0.99 | 1.00 | 0.92 | 1.00 | 1.00 | 0.99 |
| | 75% | 2.50 | 0.77 | 0.94 | 0.58 | 0.98 | 0.99 | 0.63 | 0.99 | 1.00 | 0.88 | 0.99 | 1.00 | 0.98 |
| | 100% | 0.00 | 0.59 | 0.84 | 0.34 | 0.34 | 0.97 | 0.38 | 0.92 | 0.99 | 0.42 | 0.92 | 0.99 | 0.84 |
| | 200% | -6.02 | 0.00 | 0.27 | 0.12 | 0.01 | 0.95 | 0.00 | 0.00 | 0.94 | 0.29 | 0.50 | 0.88 | 0.36 |
| 2nd Mode | 0% | ∞ | 1.00 | 1.00 | 1.00 | 1.00 | 1.00 | 1.00 | 1.00 | 1.00 | 1.00 | 1.00 | 1.00 | 1.00 |
| | 5% | 26.02 | 1.00 | 1.00 | 1.00 | 1.00 | 1.00 | 1.00 | 1.00 | 1.00 | 1.00 | 1.00 | 1.00 | 1.00 |
| | 10% | 20.00 | 1.00 | 1.00 | 1.00 | 1.00 | 1.00 | 1.00 | 1.00 | 1.00 | 1.00 | 1.00 | 1.00 | 1.00 |
| | 20% | 13.98 | 1.00 | 1.00 | 0.99 | 1.00 | 1.00 | 1.00 | 1.00 | 1.00 | 1.00 | 1.00 | 1.00 | 1.00 |
| | 50% | 6.02 | 1.00 | 1.00 | 0.99 | 1.00 | 1.00 | 1.00 | 1.00 | 1.00 | 1.00 | 1.00 | 1.00 | 1.00 |
| | 75% | 2.50 | 0.99 | 1.00 | 0.98 | 1.00 | 1.00 | 1.00 | 1.00 | 1.00 | 1.00 | 1.00 | 1.00 | 1.00 |
| | 100% | 0.00 | 0.99 | 1.00 | 0.97 | 1.00 | 1.00 | 0.99 | 1.00 | 1.00 | 0.99 | 1.00 | 1.00 | 1.00 |
| | 200% | -6.02 | 0.95 | 1.00 | 0.95 | 0.99 | 1.00 | 0.98 | 1.00 | 1.00 | 0.99 | 1.00 | 1.00 | 0.99 |
| 3rd Mode | 0% | ∞ | 1.00 | 1.00 | 1.00 | 1.00 | 1.00 | 1.00 | 1.00 | 1.00 | 1.00 | 1.00 | 1.00 | 1.00 |
| | 5% | 26.02 | 1.00 | 1.00 | 1.00 | 1.00 | 1.00 | 1.00 | 1.00 | 1.00 | 1.00 | 1.00 | 1.00 | 1.00 |
| | 10% | 20.00 | 1.00 | 1.00 | 1.00 | 1.00 | 1.00 | 1.00 | 1.00 | 1.00 | 1.00 | 1.00 | 1.00 | 1.00 |
| | 20% | 13.98 | 1.00 | 1.00 | 1.00 | 1.00 | 1.00 | 1.00 | 1.00 | 1.00 | 1.00 | 1.00 | 1.00 | 1.00 |
| | 50% | 6.02 | 1.00 | 1.00 | 1.00 | 1.00 | 1.00 | 1.00 | 1.00 | 1.00 | 1.00 | 1.00 | 1.00 | 1.00 |
| | 75% | 2.50 | 1.00 | 1.00 | 1.00 | 1.00 | 1.00 | 1.00 | 1.00 | 1.00 | 1.00 | 1.00 | 1.00 | 1.00 |
| | 100% | 0.00 | 1.00 | 1.00 | 0.99 | 1.00 | 1.00 | 1.00 | 1.00 | 1.00 | 1.00 | 1.00 | 1.00 | 1.00 |
| | 200% | -6.02 | 0.99 | 1.00 | 0.99 | 0.98 | 1.00 | 0.98 | 1.00 | 1.00 | 1.00 | 1.00 | 1.00 | 1.00 |
| 4th Mode | 0% | ∞ | 1.00 | 1.00 | 1.00 | 1.00 | 1.00 | 1.00 | 1.00 | 1.00 | 1.00 | 1.00 | 1.00 | 1.00 |
| | 5% | 26.02 | 1.00 | 1.00 | 1.00 | 1.00 | 1.00 | 1.00 | 1.00 | 1.00 | 1.00 | 1.00 | 1.00 | 1.00 |
| | 10% | 20.00 | 1.00 | 1.00 | 1.00 | 1.00 | 1.00 | 1.00 | 1.00 | 1.00 | 1.00 | 1.00 | 1.00 | 1.00 |
| | 20% | 13.98 | 1.00 | 1.00 | 1.00 | 1.00 | 1.00 | 1.00 | 1.00 | 1.00 | 1.00 | 1.00 | 1.00 | 1.00 |
| | 50% | 6.02 | 1.00 | 1.00 | 1.00 | 1.00 | 1.00 | 1.00 | 1.00 | 1.00 | 1.00 | 1.00 | 1.00 | 1.00 |
| | 75% | 2.50 | 1.00 | 1.00 | 1.00 | 1.00 | 1.00 | 1.00 | 1.00 | 1.00 | 1.00 | 0.99 | 1.00 | 0.98 |
| | 100% | 0.00 | 1.00 | 1.00 | 1.00 | 1.00 | 1.00 | 1.00 | 1.00 | 1.00 | 1.00 | 0.99 | 1.00 | 0.98 |
| | 200% | -6.02 | 1.00 | 1.00 | 0.99 | 0.99 | 1.00 | 1.00 | 0.99 | 1.00 | 0.99 | 0.97 | 0.99 | 0.96 |
| 5th Mode | 0% | ∞ | 1.00 | 1.00 | 1.00 | 1.00 | 1.00 | 1.00 | 1.00 | 1.00 | 1.00 | 1.00 | 1.00 | 1.00 |
| | 5% | 26.02 | 1.00 | 1.00 | 1.00 | 1.00 | 1.00 | 1.00 | 1.00 | 1.00 | 1.00 | 1.00 | 1.00 | 1.00 |
| | 10% | 20.00 | 1.00 | 1.00 | 1.00 | 1.00 | 1.00 | 1.00 | 1.00 | 1.00 | 1.00 | 1.00 | 1.00 | 1.00 |
| | 20% | 13.98 | 1.00 | 1.00 | 1.00 | 1.00 | 1.00 | 1.00 | 0.99 | 1.00 | 1.00 | 1.00 | 1.00 | 1.00 |
| | 50% | 6.02 | 1.00 | 1.00 | 1.00 | 1.00 | 1.00 | 1.00 | 0.96 | 1.00 | 0.99 | 1.00 | 1.00 | 1.00 |

| | | | | | | | | | | | | |
|---|---|---|---|---|---|---|---|---|---|---|---|---|
| 75% | 2.50 | 1.00 | 1.00 | 1.00 | 1.00 | 1.00 | 1.00 | 0.96 | 0.98 | 0.97 | 1.00 | 1.00 | 0.99 |
| 100% | 0.00 | 1.00 | 1.00 | 0.99 | 1.00 | 1.00 | 0.99 | 0.95 | 0.98 | 0.95 | 0.99 | 1.00 | 0.99 |
| 200% | -6.02 | 0.99 | 1.00 | 0.98 | 0.99 | 1.00 | 0.99 | 0.95 | 0.96 | 0.95 | 0.96 | 1.00 | 0.96 |

Figs. 4 to 8 show the acceptable minimum value for MAC for the first five mode shapes of beam CF identified by PP for the various noise levels together with the corresponding reference mode shapes. As PP identified the mode shapes for each node, these figures can be interpolated between two nodes. Fig. 4 shows that mode shapes up to a noise level of 20% (SNR ≥13.98) are an appropriate approximation of the first mode shape of beam CF. Figs. 5 to 8 show that the other mode shapes of this beam are in good agreement with the corresponding reference mode shapes up to a noise level of 200% (SNR ≥ -6.02). Table 10 and the figures show that 0.95 is the defined MAC threshold. It was found that this value is constant for the other methods and the other beams.

As for the mode shapes, the first natural frequency and other frequencies of beam CF were also obtained by PP at noise levels of 20% and 200%, respectively (Table 5). The mode shape was properly identified for the previously identified corresponding natural frequency. For example, the first mode shape of beam CF was identified by PP at a noise level of 20% and the corresponding natural frequency was also identified at this noise level. This was also done for the other beams and methods. The natural frequencies identified in Tables 5 to 8 have corresponding MAC values of ≥0.95 (Table 10). For example, in Tables 5 to 8, the second to fourth natural frequencies of each beam for all noise levels were identified. Their MAC values in Table 10 are all ≥0.95.

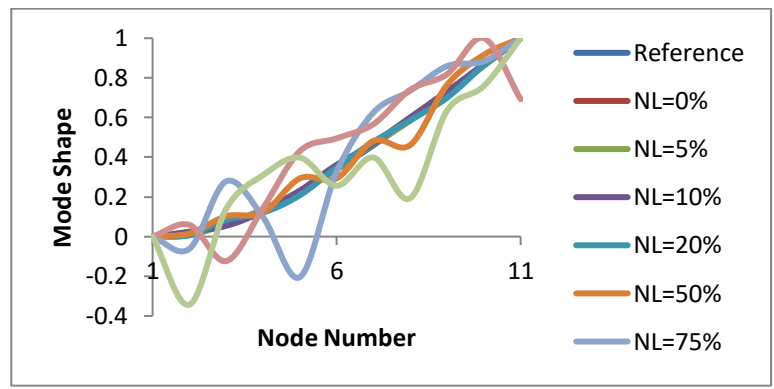

**Fig. 4.** The first mode shape of beam CF

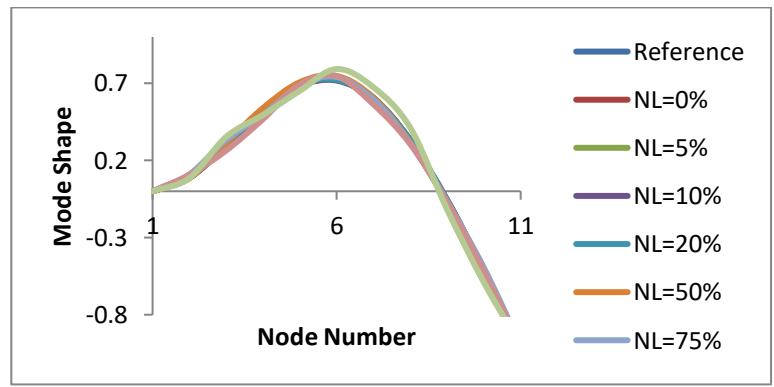

**Fig. 5.** The second mode shape of beam CF

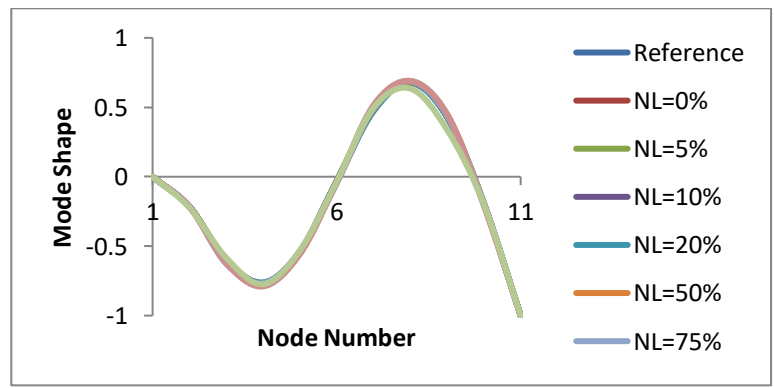

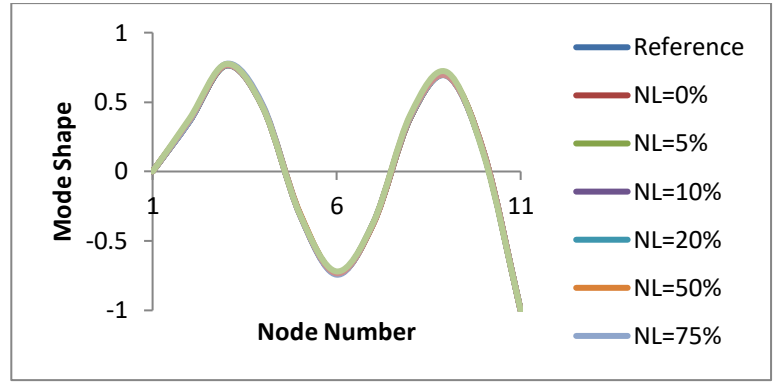

**Fig. 7.** The fourth mode shape of beam CF

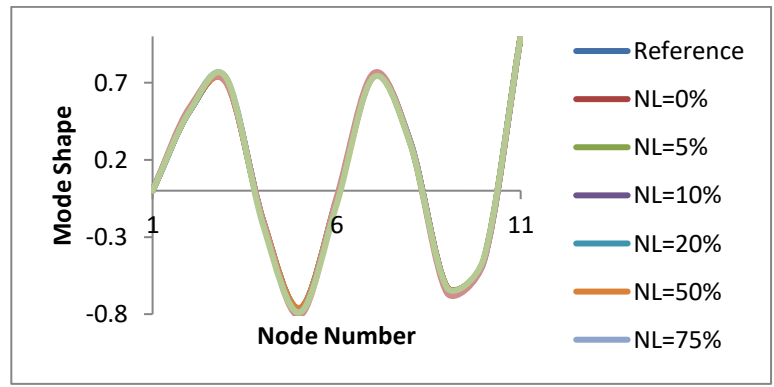

**Fig. 8.** The fifth mode shape of beam CF

## 5. Conclusion

To investigate the effect of noise on modal parameter identification, noise with different powers were added to the acceleration signals of beams and the noisy data was generated. This process was repeated 100 times. The modal parameters of the beams were identified for all 100 noisy data sets.

The natural frequencies of all modes of each beam except the first mode were identified for all noise levels. The natural frequency of the first mode was determined using the PP, FDD and SSI methods for SNR values of ≥13.98, ≥6.02 and ≥26.02, respectively. The natural frequencies of the beams identified by the PP and FDD methods remained unchanged compared to the noise-free case. FDD was more powerful for identifying the natural frequencies of the beams because it identified the natural frequencies at higher noise levels compared to the other two methods. All results obtained for identification of the natural frequencies of the beams held true for the corresponding mode shapes; that is, a mode shape was appropriately identified for every corresponding natural frequency. The MAC values between the properly identified mode shapes and the corresponding reference mode shapes were ≥0.95.

Modal parameters of the first few modes of a structure are used for structural health monitoring, finite element model updating and damage detection. The results of this study have shown that the accuracy of the first five modes identified vary according to the amount of noise in the data. Because the amount of noise is unknown in practice, the parameters of the higher modes identified using output-only methods are more reliable than for the first mode of the beams. It is important to note that these results hold true for this particular structural system, mesh length and noise model generation. However, further studies have shown that changing the reference channel, beam section height, recording duration, sampling time, response of beam and RMS amplitudes of input excitation does not change the results.